\documentclass[final,pre,twocolumn,superscriptaddress,floatfix]{revtex4-1}
\usepackage[english]{babel}
\usepackage{latexsym}
\usepackage{amsmath}
\usepackage{color,soul}
\usepackage{hyperref}
\usepackage{fancyhdr}
\usepackage{graphicx}
\usepackage{wrapfig}
\usepackage{hhline}
\usepackage{dcolumn}
\usepackage[english]{babel}
\usepackage{amsmath}
\usepackage{amsfonts}
\usepackage{graphicx}
\usepackage{xprintlen}

\usepackage{microtype}

\usepackage{latexsym}
\usepackage{times}
\usepackage{amsmath}
\usepackage{color,soul}
\usepackage{fancyhdr}
\usepackage{wrapfig}
\usepackage{hhline}
\usepackage{dcolumn}
\usepackage[normalem]{ulem} 

\graphicspath{{./}{./images/}}

\usepackage{color}

\newcommand{\beq}{\begin{equation}}
\newcommand{\eeq}{\end{equation}}

\begin{document}

\title{A network model of conviction-driven social segregation}
%
%
%
%

\author{Gianluca Teza}
\email{gianluca.teza@phd.unipd.it}
\affiliation{Dipartimento di Fisica e Astronomia G. Galilei, University of Padova, Via Marzolo 8, Padova, Italy}

\author{Samir Suweis}
\affiliation{Dipartimento di Fisica e Astronomia G. Galilei,
  University of Padova, Via Marzolo 8, Padova, Italy}

\author{Marco Gherardi}
\affiliation{Sorbonne Universit\'e, UPMC Univ Paris 06, UMR 7238,
  Computational and Quantitative Biology, 4 Place Jussieu, Paris, France}
\affiliation{Current address: Dipartimento di Fisica, Universit\`a
  degli Studi di Milano, via Celoria 16, 20133 Milano, Italy}

\author{Amos Maritan}
\affiliation{Dipartimento di Fisica e Astronomia G. Galilei,
  University of Padova, Via Marzolo 8, Padova, Italy}

\author{Marco {Cosentino Lagomarsino}}
\email{marco.cosentino-lagomarsino@upmc.fr}
\affiliation{Sorbonne Universit\'e, UPMC Univ Paris 06, UMR 7238,
  Computational and Quantitative Biology, 4 Place Jussieu, Paris, France}
\affiliation{CNRS, UMR 7238, Paris, France}
\affiliation{IFOM, FIRC Institute for Molecular Oncology, Milan, Italy}

\begin{abstract}
  In order to measure, predict, and prevent social segregation, it is
  necessary to understand the factors that cause it. While in most
  available descriptions space plays an essential role, one
  outstanding question is whether and how this phenomenon is possible
  in a well-mixed social network.  We define and solve a simple model
  of segregation on networks based on discrete convictions. In our
  model, space does not play a role, and individuals never change their
  conviction, but they may choose to connect socially to other
  individuals based on two criteria: sharing the same conviction, and
  individual popularity (regardless of conviction). The trade-off
  between these two moves defines a parameter, analogous to the
  ``tolerance'' parameter in classical models of spatial segregation.
  We show numerically and analytically that this parameter determines
  a true phase transition (somewhat reminiscent of phase separation in
  a binary mixture) between a well-mixed and a segregated state.
  Additionally, minority convictions segregate faster and
  inter-specific aversion alone may lead to a segregation threshold
  with similar properties. Together, our results highlight the general
  principle that a segregation transition is possible in absence of
  spatial degrees of freedom, provided that conviction-based rewiring
  occurs on the same time scale of popularity rewirings.
\end{abstract}

\maketitle

\section{Introduction}

%
Social segregation is a primary problem for our well-being, and for
the policy-making of our governments. The most basic questions
regarding social segregation concern its quantification, and the
prediction and prevention of its onset and its outcomes. Attempts to
approach the problem from a quantitative viewpoint date back to the
late 1960s, with a model proposed by the economist Thomas
C.~Schelling~\cite{Schelling1971,Schelling1969}. In this model,
individuals are embedded in a two-dimensional lattice, and are
characterized by a threshold ``tolerance'' to other individual
opinions.  This model naturally attracted the attention of statistical
physics because of its analogy with Blume-Emery-Griffiths and Potts
models, and more in general with binary mixtures and interfacial
dynamics.
It shows a complex phase diagram, including threshold phenomena (phase
transitions) where opinions separate spatially and may form
patterns~\cite{DallAsta2008,Gauvin2010,Rogers2012,Gauvin2009}.
Schelling's model demonstrates that even mild preferences for a set of
agents for defining themselves as a local minority can produce strong
spatial segregation patterns, challenging the common view that
discrimination is a necessary condition for segregation.

%
While spatial ``steric'' interactions and dimensionality are very
important in Schelling's model, human interactions can in most cases
be described as
network-like~\cite{Newman2002,Watts1998,Amaral2000,Barthelemy2003a,Barthelemy2011}.
In a situation with (nearly) immutable convictions and limited
tolerance to other opinions, individuals sharing the same conviction
might find themselves severed from society even if their potential for
social interaction is not limited by spatial constraints. Such a
situation is very dangerous for society, for the danger of triggering
self-propelled distortions of reality shared between many individuals.
For example, this is particularly relevant in the on-line world of
social networks.  The diffusion of on-line non-intermediated
unverified and polarized contents and the spread of misinformation is
becoming a pressing problem for our society. One of the most relevant
driving forces has been recognised as the echo-chamber
effect~\cite{Sirbu2013,Zollo2015,DelVicario2016a}. It consists in the
formation of segregated clusters of users who share some strong common
opinions, increasingly reinforcing these ideas and thus becoming
impenetrable to news diverging from their point of view.

Thus, another possible approach (relatively less explored) may attempt
to describe segregation using opinion-based network models, such as
the voter model~\cite{Castellano2009,Sood2005,Suweis2012}. The complex
networks literature provides many examples of segregation in the
structure of relationships (from school friendship to value- and
belief-oriented partitioning) empirical
data~\cite{Girvan2002,Newman2004}. However, the literature on complex
networks models focuses mostly on how opinion dynamics is shaped by
network-like human interactions, i.e., on how individuals change their
mind based the opinions of others~\cite{Ben1996,Sood2005,Suweis2012}.
Such a framework is not well-suited to describe segregation, where
precisely the opposite occurs, i.e., human interactions change
following stable ``opinions'', or other more general
individual-specific factors (as it happens in Schelling's model).
Indeed, some of these factors may be very strongly rooted in
individuals, such as convictions, religious and cultural factors, and
even immutable physical or racial features.
A comparativelly smaller thread of
studies~\cite{Holme2006,Castellano2009,Durrett2012,Min2017} has
considered the coevolution of network connections and opinions. In
such models, individuals can both change their mind and change their
connections, and segregated states can emerge, depending on the
intrinsic time scales of these
processes~\cite{Holme2006,Durrett2012}. However, the conditions for
reaching segregated states are not the main focus of these
investigations, which are typically focused on the conditions for
reaching consensus. In order to understand the factors leading to
segregated states, it is important to address the case where node
attributes (convictions) are persistent.

%
There is very little work in the literature addressing such situation
on networks.  A fairly recent study~\cite{Henry2011}, considered the
emergence of segregation in a social network by a model with
continuous opinions and an individual ``aversion bias'' favoring the
severing of connections with increasing difference of opinions, in
favor of random rewiring. They proved the existence of attractor
steady states with given segregation levels that are independent of
initial conditions, and characterized the time scales of convergence
to these states.
However, this study did not address the possibility and existence of
the threshold phenomena that are ubiquitious in Schelling's
model. Such phenomena are important to address, as argued in the
previous paragraphs.

Here, we define an alternative model of segregation on networks based
on \emph{discrete} convictions, and we study it through analytical
calculations and direct simulation. In our model, individuals may
choose to follow other individuals based on sharing the same
conviction, or based on their popularity (regardless of
conviction). The trade-off between these two moves defines a
transition between a well-mixed and a segregated state. A threshold
parameter, analogous (but not equivalent) to the ``tolerance''
parameter in Schelling's model, weighs the two different possible
choices.  We analyze this model in the case of binary states of the
agents (two possible convictions, such as Democrats and Republicans),
and we are able to fully characterize the conditions for the emergence
of phase transitions the relaxation time scales of the system in the
segregated and non-segregated phases. Importantly, in order for
transitions to exist, the conviction move has to occur on the same
time scale of the popularity move, regardless of the size of the
community being segregated.  Finally, we show that minority
convictions segregate more easily, and we characterize this phenomenon
quantitatively.
%
%

\begin{figure}
\centering
\includegraphics[width=0.38\textwidth]{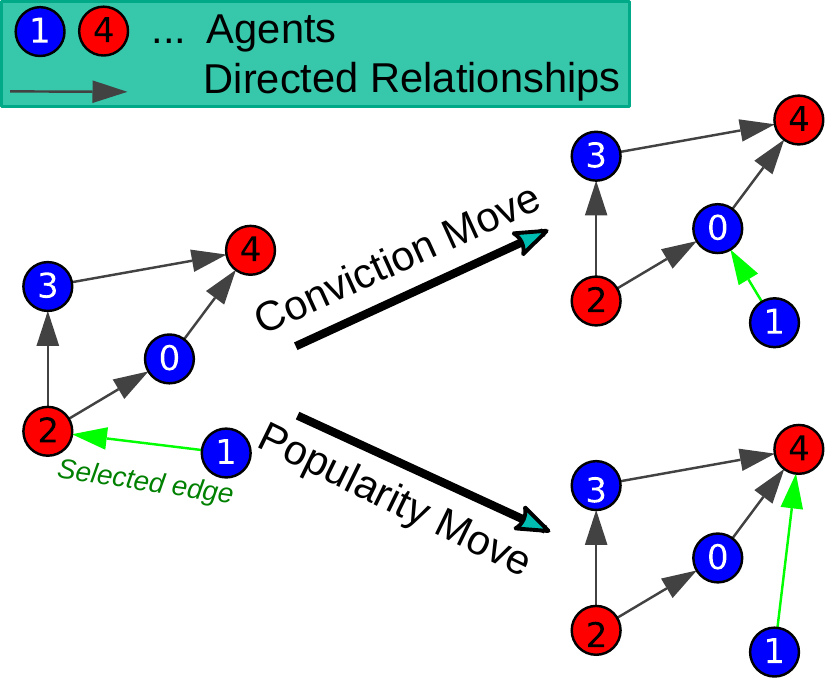}
\caption{\label{figure:moves} Illustration of the action of the model
  basic moves. Nodes represent agents and colors represent
  convictions.  Edges represent directed social connections (A follows
  B if an edge is sent from A to B). The selected edge to be removed
  is in both cases $e_{1\to 2}$. In a conviction move, the new target
  can be chosen only among the blue nodes (in the sketch this move
  creates the edge $e_{1\to 0}$), while in a popularity move the new
  target can be chosen regardless of its opinion, so that every node
  with an in-degree greater than 0 is a potential candidate (in the
  sketch this moves creates the edge $e_{1\to 4}$).}
\end{figure}

\section{Definition of the model}
Our model describes a social network as a directed graph where
individuals (nodes) follow other individual's opinions by sending
directed edges to their corresponding nodes.  The initial condition is
a random directed graph $G_0(N,m,h)$ made of $N\in\mathbb{N}$
nodes. Each node has fixed outdegree $m\in\mathbb{N}$. A fraction
$h\in[0;1]$ of individuals hold a certain conviction, which we
identify with the color \emph{red} (as opposed to the probability
$1-h$ of holding the opposite conviction, i.e. being colored in
\emph{blue}). The total number of edges $M=N\cdot m$ defines the size
of our system.  The graph is constructed through the associated
adjacency matrix by filling randomly with $m$ ones the matrix rows of
a zero matrix (we exclude the matrix diagonal elements which would
indicate self-edges). As a consequence of this construction procedure,
the in-degrees follow a Poisson distribution with average value $m$ (as
in an Erd\~os-R\'enyi random graph \cite{Erdos1960}).

The network evolves at \emph{fixed} conviction, by choosing at each
step one of two possible rewiring moves (Fig.~\ref{figure:moves})
accordingly to the choice parameter $\varphi\in[0;1]$
:
\begin{itemize}
\item with probability $\varphi$ a \emph{conviction move} chooses
  randomly one among all the edges $e_{i\to j}$ between two nodes
  holding different convictions (which we will call ``heterogeneous'' edges),
  deletes, chooses uniformly a new target node $k$ holding the same
  conviction as $i$ and creates a new ``homogeneous'' edge $e_{i\to k}$;
\item alternatively, with probability $1-\varphi$, a \emph{popularity
    move} which chooses randomly one edge $e_{i\to j}$ among all the
  edges of the network, deletes it, and creates a new edge
  $e_{i\to k}$ with a target $k$ chosen among all the nodes with a
  preferential attachment criterion, i.e. with a probability equal to
  the in-degree of the target node normalized by the total number of
  edges $M$.
\end{itemize}
It is important to underline the fact that the opinion move selects
the edge to be removed in the basket of the heterogeneous edges.  As it
will be more clear in the following, this choice is essential in order
to obtain a threshold phenomenon  for segregation.

We quantify the segregation using as order parameter the total number
of homogeneous edges connecting nodes with the same conviction. In the
initial condition ($t=0$), and for $M$ sufficiently large, the
densities of the four different kinds of edges (red to red, blue to
blue, red to blue and blue to red) are:
\begin{align}
e_0(rr) &=h^2 \nonumber\\
e_0(bb) &=(1-h)^2 \nonumber\\
e_0(rb) &=e_0(br)=h(1-h) \ \ . 
\end{align}
More in general, for every step $t>0$, the link densities are
functions of this parameter order parameter. Indeed, since
$\Omega_t:=M(e_t(rr)+e_t(bb))$, one has
\begin{align}
e_t(rr) &=\frac{h^2}{h^2+(1-h)^2} \frac{\Omega_t}{M} \nonumber\\
e_t(bb) &=\frac{(1-h)^2}{h^2+(1-h)^2} \frac{\Omega_t}{M} \nonumber\\
e_t(rb) &=e_0(br)=\frac{M-\Omega_t}{2M} \ . 
\end{align}
We define a segregated phase as a state where, for large networks,
typically all the heterogeneous edges disappear, leaving the network
with only edges between like-minded nodes, characterized by a
saturation of the order parameter to the maximum value $\Omega_t=M$.

\section{Results}

\begin{figure*}
\centering
\includegraphics[width=0.6\textwidth]{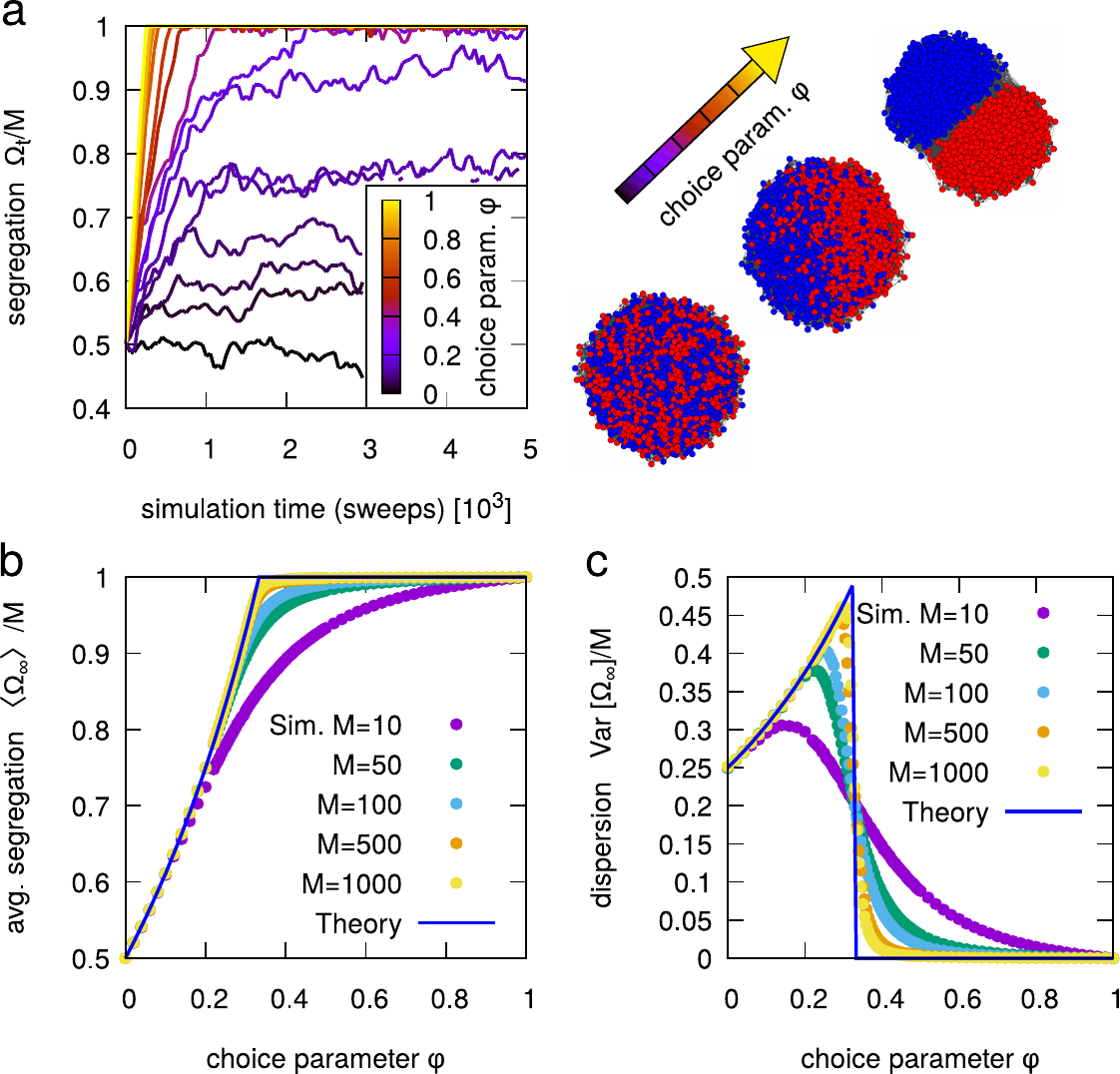}
\caption{\label{figure:ph_transition} A threshold phenomenon to a
  segregated state appears for a critical value of the choice
  parameter $\varphi_c$.  \textbf{a)} Evolution of the fraction of
  homogeneous links.  The plot shows the order parameter normalized by
  the total number of edges $M$ plotted against sweeps. The curves are
  obtained by simulating the evolution of the same initial random
  graph $G_0(N=500,m=5,h=1/2)$ for different values of $\varphi$. For
  low $\varphi$, the long-time value of $\Omega_{\infty}(\varphi)$
  relaxes to a steady state where the edges connecting nodes with
  different colors fluctuate around a finite value, while as $\varphi$
  grows, it reaches one (a segragated state) in a finite time. The
  right-hand panel shows some illustrative simulation snapshots, where
  the network is visualized with a spring model based on shared links.
  \textbf{b)} Plot of the mean order parameter at steady state versus
  the choice parameter $\varphi$ comparing the analytical results
  (solid line) of Eq.~\ref{eq:Omega_SS_phi} with numerical simulations
  for different sizes of the network $M$ (symbols).  This analysis
  supports a segregation transition for $\varphi_c=1/3$ (for $h=1/2$).
  \textbf{c)} Fluctuations scale linearly with the size of the
  system. Plot of the dispersion of the order parameter from the
  simulations in panel b (symbols). As the size of the network grows,
  the variability across realizations peaks around the critical value
  $\varphi_c=1/3$ reflecting the prediction of
  Eq.~\ref{eq:Omega2_SS_phi} (solid line).}
\end{figure*}

\subsection{A transition to a segregated state emerges at a critical
  point}

By construction of the model dynamics, conviction moves favor the
transition to a segregated phase, while popularity moves try to
reestablish the disorder and will also affect the in-degree
distribution. Moreover, we expect networks characterized by asymmetric
densities of opinions ($h\neq 1/2$) to reach a segregated phase more
easily.

Starting by the same initial random graph $G_0$, we evolved the
network for different values of $\varphi$ and at each step we recorded
the order parameter $\Omega_t(\varphi)$, starting from initial
conditions with $\Omega_0=1/2$ for $h=1/2$
(Fig.~\ref{figure:ph_transition}a), representing the fraction of
homogeneous edges (connecting individuals with equal convictions). For
low values of $\varphi$, the system does not segregate, but they reach
a balance between popularity- and conviction-based moves. As the value
of $\varphi$ increases, conviction-based moves become increasingly
dominant, and the steady-state value of the order parameter increases
until it reaches the maximum possible value $M$, indicating that
typically the number of heterogeneous edges is negligible compared to
the total number of edges, and the system reaches a segregated
phase. This behavior suggests the existence of a critical value
$\varphi_c$ of the choice parameter, above which the steady state of
the network is always in a segregated phase.

In order to find the critical value of the choice parameter
analytically, we used a mean-field approach, based on an estimate of
the average variation $\Delta\Omega_t$ at every step. Conviction moves
increase $\Omega_t$ by 1, while popularity moves might act differently
depending on the probability of picking an edge of a certain kind, and
also on the kind of the new edge created.  The resulting mean-field equation is 
\begin{equation}
  \Delta\left<\Omega_{t}(\varphi,h)\right> =
  \underbrace{\vartheta\varphi}_{\text{conv. move}} +
  \underbrace{(1-\varphi)\left[\vartheta
      p_{t}^{+}(h)-p_{t}^{-}(h)\right]}_{\text{pop. move}}   \ ,
\label{eq:midfield}
\end{equation}
where the Heaviside step function $\vartheta:=\theta(M-\Omega_t)$
excludes  forbidden moves once the segregation state is
reached, while $p_t^{\pm}(h)$ are the probabilities of respectively
increasing and decreasing the order parameter with a popularity move. 

In the continuum time limit, and for $h=1/2$ (for a more general
derivation for every $h\in[0;1]$ see section \ref{subsec:meanfield})
Eq.~\eqref{eq:midfield} gives the following differential equation for the
average value of the order parameter
\begin{equation}
\partial_{t}\left< \Omega_{t} (\varphi) \right>
=\vartheta\frac{1+\varphi}{2}-(1-\varphi)\frac{1+\vartheta}{2}\frac{\left<
    \Omega_{t}(\varphi)\right>}{M}  \ .
\label{eq:dt_Omega}
\end{equation}
This equation can be explicitly integrated (for $\varphi\neq1$),
yielding the  time dependence for the average value of the
order parameter, 
\begin{align}\label{eq:Omega_time}
  \frac{\left< \Omega_{t} (\varphi) \right>}{M}
  &=
    \left[ \left( 1-\frac{1}{2}\vartheta\right) -
    \frac{\vartheta}{1+\vartheta}\frac{1+\varphi}{1-\varphi}\right]
    e^{-(1-\varphi)\frac{1+\vartheta}{2M}t} +
    \nonumber\\   
  &+\frac{\vartheta}{1+\vartheta}\frac{1+\varphi}{1-\varphi}\ \ . 
\end{align}
In the pre-segregation regime (where $\Omega_t<M$ and therefore
$\vartheta=1$) the relaxation is then exponential  with
characteristic time
\begin{equation}
\tau_{\Omega}=\frac{M}{1-\varphi}.
\label{eq:tau_Omega}
\end{equation}
Hence, the asymptotic value
\begin{equation}
\frac{\left< \Omega_{\infty} (\varphi)
  \right>}{M}=\min_{\varphi\in[0;1)} \left\{
  1,\frac{1+\varphi}{2(1-\varphi)}\right\} \label{eq:Omega_SS_phi} 
\end{equation}
will be reached for times
$t\gg\tau_{\Omega}$. Fig.~\ref{figure:ph_transition}b compares this
prediction with direct simulations. The model behaves as expected
already for relatively small-sized networks ($M=100$) and gradually
moves towards the predicted curve as the size of the system grows. By
setting $\left< \Omega_{\infty} (\varphi) \right>=1$ in
Eq.~\ref{eq:Omega_SS_phi} and solving for $\varphi$ one finds the
critical value of the choice parameter at which the transition occurs,
which for $h=1/2$ is $\varphi_c=1/3$.  This transition has a clear
similarity with second order phase transitions~\cite{Landau1980} ,
because of a discontinuity in the first derivative of $\Omega_t$ with
respect to $\varphi$. The analogy identifies the order parameter
$\Omega$ with the magnetization, while the role of the temperature is
played here by the choice parameter $\varphi$.

The fluctuations of the order parameter also characterize the
transition. These can be estimated by the second cumulant moment
$\text{Var}[\Omega_{\infty}(\varphi)]$. A peak in amplitude of the
fluctuations at the critical value $\varphi_c$ should signal the
transition. In the social segregation interpretation, this means that
the transition to a segregated state is also marked by sudden growth
and shrinkage of its connections to the rest of the world. In order to
access the fluctuations analytically, we explicitly considered the
master equation~\cite{Gardiner1985}). Calling $P_t(\Omega)$ the
probability of having $\Omega$ homogeneous edges at  time $t$ the master
equation is defined as 
\begin{equation}
\partial_t P_t (\Omega)=\sum_{\Omega'\neq\Omega}
W(\Omega|\Omega')P_t(\Omega')-
W(\Omega'|\Omega)P_t(\Omega) \ ,
\label{eq:ME}
\end{equation}
where $W(\Omega|\Omega')$ are the transition rates of moving from a
network with $\Omega'$ homogeneous edges to a network of $\Omega$ edges, which
for our system (always in the case of $h=1/2$) is
\begin{align}\label{eq:ME_rates}
  W(\Omega|\Omega')
  &=\delta_{\Omega',\Omega-1}\left[
    \varphi+(1-\varphi)\frac{M-\Omega'}{2M} \right] +
    \nonumber\\ 
  &+\delta_{\Omega',\Omega+1}(1-\varphi)\frac{\Omega'}{2M}
    +\delta_{\Omega',\Omega}\frac{1-\varphi}{2} \ . 
\end{align}
In the above equation, the first row describes the contribution of
both the opinion and popularity moves to an increase in $\Omega$,
while the second row describes the contributions of the popularity
move to respectively decrease and keep unaltered the order
parameter. Then we define the factorial moment generating function
\begin{equation}
  G(s,t)=\sum_{\Omega=0}^{M}s^{\Omega}P_t(\Omega) \ ,
  \label{eq:FGMF}
\end{equation}
where $s\in\mathbb{R}$ is the dual parameter of $\Omega$. Combining
Eqs.~\eqref{eq:ME} and~\eqref{eq:FGMF} (see
Appendix~\ref{subsec:ME_FMGF}) yields the following partial
differential equation,
\begin{equation}\label{eq:FGMF_dyn}
\partial_t G(s,t)=G(s,t)\frac{1+\varphi}{2}\left( s-1 \right)+
\partial_s G(s,t)\frac{1-\varphi}{2M}\left( 1-s^2 \right) \ . 
\end{equation}
By evaluating $\partial_s^n[\partial_t G(s,t)|_{s=1}]$ for every
$n\in\mathbb{N}$ we obtain a closed system of time-only differential
equations giving the exact dynamics (including the transient phase) of
all the factorial moments. The first factorial moment coincides with
the average, so we find again Eq.~\ref{eq:dt_Omega}, whereas the
second factorial moment gives $\left < \Omega_t^2 \right > $ and hence
the variance.  Taking the long-time limit we obtain an analytical
expression for the fluctuations
\begin{equation}
\frac{\text{Var} [\Omega_{\infty}
  (\varphi)]}{M}=
  \begin{cases}
  \frac{1+\varphi}{4(1-\varphi)} & \text{for } \varphi\leq1/3\\
  0 & \text{for } \varphi>1/3
  \end{cases}
\label{eq:Omega2_SS_phi} 
\end{equation}
Fig.~\ref{figure:ph_transition}c shows that as the size of $M$ (number
of edges) of the network grows, the simulations tend to agree with
this large-$M$ prediction, showing a behavior that resembles that of
the susceptibility in second-order phase transitions, with
fluctuations amplitude scaling linearly in $M$.

\begin{figure}
\centering
\includegraphics[width=0.48\textwidth]{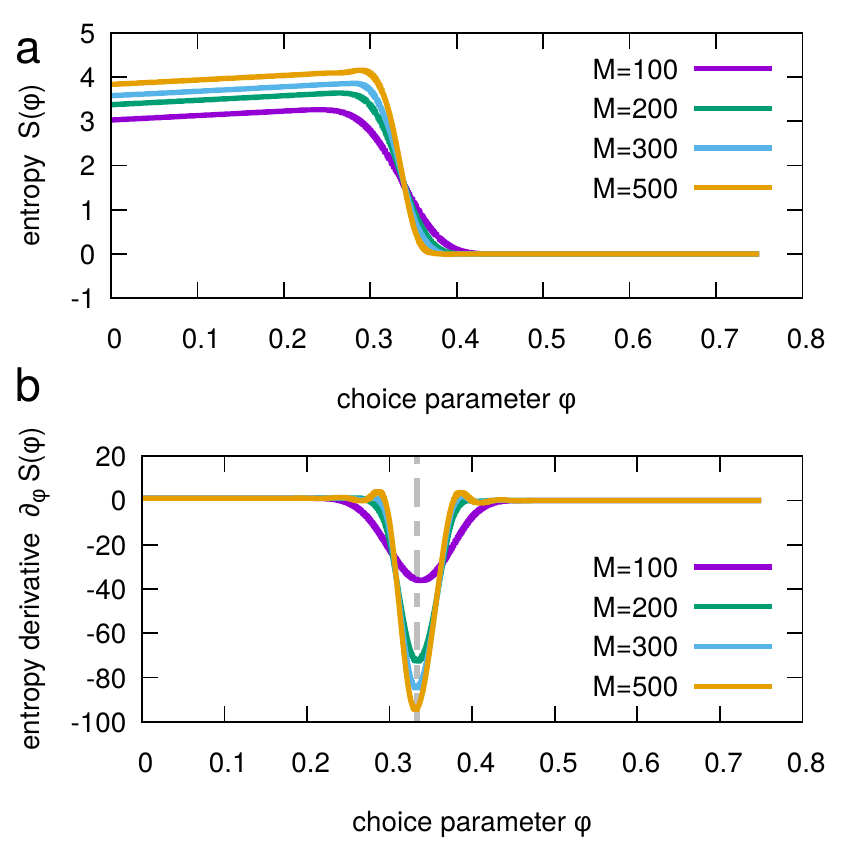}
\caption{\label{figure:entropy-phase-transition} Entropy is
  characterized by a discontinuity in correspondence with the critical
  value of the choice parameter $\varphi_c$. A) The entropy of the
  system as a function of the order parameter $\varphi$ for different
  system size $M$. B) Its derivative with respect to $\varphi_c$. The
  dashed gray line represent the predicted critical threshold
  $\varphi_c=1/3$.}
\end{figure}

By means of the generating function formalism, we can go further and
calculate exactly the stationary solution of the Master Equation
(\ref{eq:ME}) with transition rates given by
Eq. (\ref{eq:ME_rates}). The resulting stationary probability function
$P_{\text{stat}}$ is (see Appendix~\ref{subsec:ME_STATSOL} for
detailed calculations):
\begin{equation}\label{Pstaz}
  P_{\text{stat}}(\Omega )=
  \frac{2^{-\frac{M (\varphi +1)}{1-\varphi}}
    \left(\frac{M (\varphi+1)}{1-\varphi }\right)^{(\Omega)}}
  {\Omega !},  
\end{equation}
where $x^{(\Omega )}$ is the factorial power of $x$ and it is given by
$\frac{\Gamma (x+1)}{\Gamma (-\Omega +x+1)}$.  From Eq.~(\ref{Pstaz})
we can then define the entropy of the system
$S(\varphi)=-\sum_{\Omega=0}^{M\rightarrow\infty}
P_{\text{stat}}(\Omega) \log[P_{\text{stat}}(\Omega )]$ and its
derivative with respect to the choice parameter $\varphi$. As Figure
\ref{figure:entropy-phase-transition} shows, by plotting $S(\varphi)$
and $\partial_{\varphi}S(\varphi)$ we can effectively see that the
system undergoes a genuine phase transition.

\subsection{Overlap of time scales is necessary for a segregation
  transition to exist}

We now discuss more in detail an essential ingredient for a
segregation sharp transition to exist, the fact that the conviction
move occurs on the same time scale of the popularity move, regardless
of the size of heteorogeneous edges in the system. In other words, the
conviction move is realized at each step with probability $\varphi$
drawing directly from the basket of heterogeneous edges in order to
observe the transition.

We can understand this result by considering a similar model in which
the opinion move is, for instance, defined as follows. Select an edge
randomly among all the $M$ edges of the network (rather then from the
basket of the heterogeneous ones) and if the edge is heterogeneous
execute the conviction move, otherwise leave the network unaltered and
move on by executing a new step.  In this model the mean-field
equation, Eq.~\ref{eq:midfield} will take an additional term
representing the heterogeneous edge density multiplying the conviction
move term, 
\begin{equation}
\Delta\left<\Omega_{t}\right>=\underbrace{\vartheta\varphi\frac{M-\left<\Omega_t\right>}{M}}_{\text{op. move
    variant}}+(1-\varphi)\left[\vartheta
  p_{t}^{+}(h)-p_{t}^{-}(h)\right] \ . 
\label{eq:midfield_variant}
\end{equation}
The critical value $\varphi_c$ is found setting
$\Delta\left<\Omega_{t}\right>$ to zero and the average value of the
order parameter saturates to its maximum value $M$. Substituting these
quantities one immediately finds that the contribution of the opinion
move disappears, leaving us with the equation
$(1-\varphi_c)\left[\vartheta p_{t}^{+}(h)-p_{t}^{-}(h)\right]=0$
which has the only trivial solution $\varphi_c=1$ (that represents a
model in which only opinion based move are executed). In other words,
a segregated phase is found only in the trivial case where the agents
only choose their connections by conviction.

%
%
This analysis also gives a general condition for the existence of a
transition, which is that the conviction move has to be such that the
multiplicative factor introduced in the opinion move term in
Eq.~\eqref{eq:midfield_variant} translates into a function
$f(\Omega_t)$ characterized by the condition $f(M)\neq0$).
A possible justification for this forcing in the opinion move can be
found by considering some realistic situations characterized by a
segregation phenomenon driven by strong convictions (ethnicity,
political orientation, religious beliefs, etc.). If an agent is left
only with opposite minded neighbors, it is likely going to be the
first one to decide to sever a connection and rewire with someone with
the same conviction. For this reason, we believe that direct targeting
of heterogeneous connection in an environment of strong convictions
might be a realistic assumption.

\subsection{The popularity move broadens the in-degree distribution in
  the unsegregated phase, but does not affect the transition point. }

\begin{figure}
\centering
\includegraphics[width=0.48\textwidth]{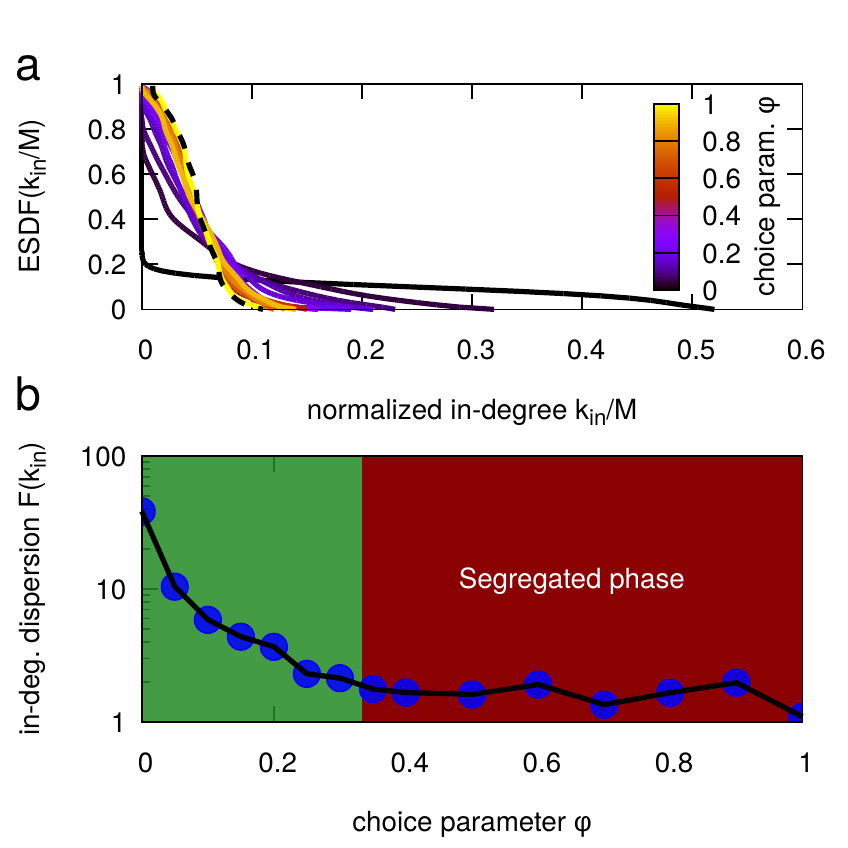}
\caption{\label{figure:in-degrees} Preferential attachment from the
  popularity move broadens hte in-degree distribution.  \textbf{a)}
  Empirical survival distribution function (ESDF) of the in-degree
  distributions of networks evolved for different values of
  $\varphi$. The plot was obtained by evolving an initial random graph
  $G_0(N=100,m=5,h=1/2)$ for $t=10^6$ steps (the in-degrees are
  normalized with respect to the total number of edges $M=500$). The
  broadening of the distribution indicates the increasing presence of
  bigger attractors in the evolved networks.  \textbf{b)} Two
  different trends for the Fano factor of the in-degrees are observed
  in the regions below and above the segregation transition. The plot
  reports the Fano factor of the in-degrees distributions shown in
  panel a versus the choice parameter $\varphi$. In the region above
  the critical value of the choice parameter $\varphi_c=1/3$ the
  deviation from a Poisson distribution ($F(k_{in})=1$) is 
  small, while the unsegregated region shows a
  super-exponential departure (the vertical axis is in log-scale)
  towards  larger dispersions as $\varphi$ decreases.}
\end{figure}

We proceed by considering the role of the popularity move in setting
the in-degree distribution and in the segregation transition. The
initial random graph $G_0(N,m,h)$ has by definition
Poisson-distributed in-degrees $k_{in}$ for large $N$,
with a mean equal do the fixed outdegree of every node of the network
$m$.  As the network evolves, the distribution of the in-degrees
changes at each popularity move, because the most popular nodes are
more likely to be chosen as a target for the newly created edges. This
determines a departure from the initial distribution towards
heavier-tailed distributions, in analogy with the ``rich gets richer''
principle that usually characterizes social
networks~\cite{Castellano2009}.
In order to properly characterize this behavior evaluated the
empirical survival distribution function (ESDF) of the in-degree
distributions of evolved graphs $G_t$ for different values of the
choice parameter. The ESDF indicates the probability of observing a
node $i$ with in-degree $k_{in}(i)$ greater then a certain value
$k_{in}$, and is defined as
\begin{equation}
\text{ESDF}(k_{in})=\frac{1}{M}\sum_{i=0}^{M}\theta(k_{in}-k_{in}(i))
\ ,
\end{equation}
Fig.~\ref{figure:in-degrees}a shows that when $\varphi=1$ the initial
distribution is unaltered (the dashed line represents the distribution
for the initial random graph $G_0$), but as $\varphi$ decreases the
in-degree distributions take increasingly heavier tails.

The same phenomenon can be quantified by a single broadness parameter
such as the Fano factor of the in-degrees $F(k_{in})$, defined as
\begin{equation}
F(k_{in})=\frac{\text{Var}[k_{in}]}{\left< k_{in} \right>} \ . 
\end{equation}
This parameter is 1 for a Poisson distribution, whereas greater values
indicate larger dispersion. Fig.~\ref{figure:in-degrees} shows this
parameter plotted as a function of the choice parameter $\varphi$. The
Fano Factor increases as popularity-based moves become more probable
(as $\varphi$ goes to zero). Moreover two different trends appear to
characterize the region below and above the critical value
$\varphi_c=1/3$.

\begin{figure}
\centering
\includegraphics[width=0.48\textwidth]{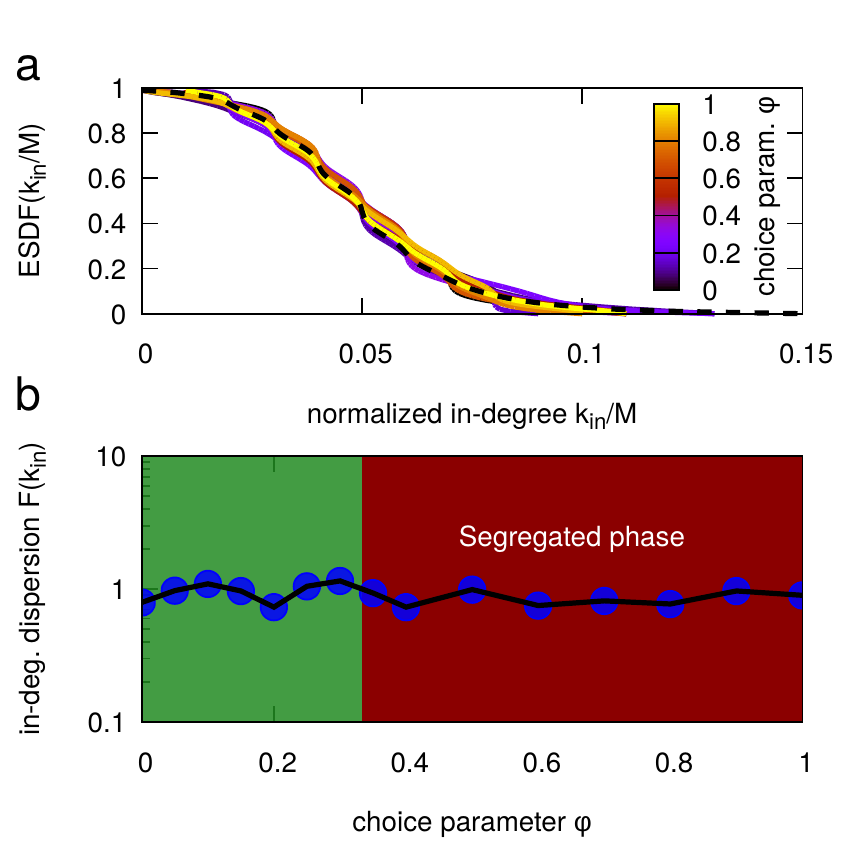}
\caption{\label{figure:in-degrees_no_pop}. Skewed node popularity does
  not affect segregation.  \textbf{ab)} Same plots as
  Fig.~\ref{figure:in-degrees}, for a model in which the popularity
  move is changed with a rewiring on a uniformly chosen random
  node. This model shows the same phase transition as the original one
  (and in particular the plots in Fig.~\ref{figure:ph_transition}bc
  are identical), but the transition is not accompanied by changes in
  node degree.}
\end{figure}

Finally, although we found that popularity-based rewiring increases
the dispersion of social connections in the unsegregated regime, this
preferential attachment ingredient does not affect the segregation
transition in any way, as we have verified by substituting
popularity-based rewiring with random rewiring in our simulations
(Fig.~\ref{figure:in-degrees_no_pop}).
Although one may expect that the presence of popular individuals may
help avoiding the emergence of segregation due to their capacity of
attracting new nodes regardless of their opinion, this does not happen
in this model. The reason is easily understood from
Eq.~\eqref{eq:midfield} and \eqref{eq:midfield_variant}, which govern
the dynamics of the order parameter, where it is clear that the
in-degree distribution never comes into play.

\subsection{Minority convictions segregate more easily}
\label{sec:min_conv}

\begin{figure}
\centering
\includegraphics[]{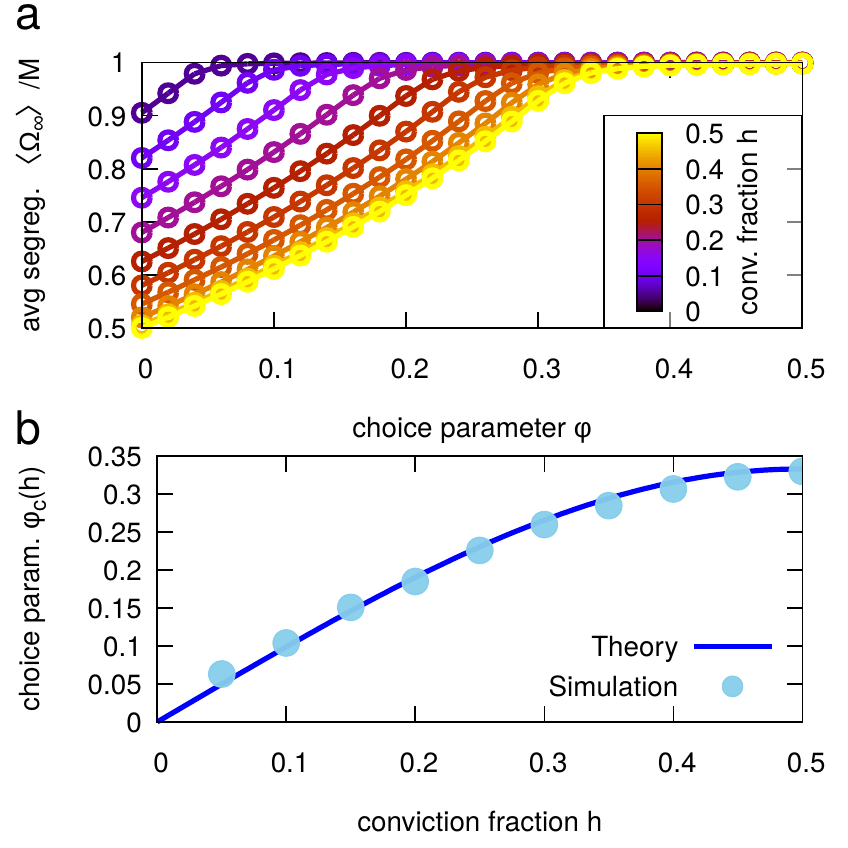}
\caption{\label{figure:critical_h} Minority convictions tend to
  segregate more easily. \textbf{a)} Average value of the order
  parameter $\left< \Omega_t \right>$ in networks evolved from initial
  networks $G_0(100,5,h)$ for different values of $h\leq1/2$ (the
  results for $h>1/2$ are the same due to the symmetry $h\to1-h$). As
  the the density of nodes holding a certain conviction decreases, the
  networks will reach a segregated phase for lower values of
  $\varphi$.  \textbf{b)} Simulations confirm the analytical
  prediction for the critical points of the model. The critical points
  (symbols) are ectracted from the curves in panel a, for different
  values of $h$, and compared with the prediction described by
  Eq.~\ref{eq:critical_h} (solid line).}
\end{figure}

The results presented up to this point were obtained under the
hypothesis of equally represented convictions condition ($h=1/2$). A
more generic case describes minority versus majority convictions,
characterized by different values of $h$.  The differences from the
symmetric case concern both the characteristic time $\tau_{\Omega}$
needed to reach the steady state and the critical value $\varphi_c$ at
which the transition to a segregated phase occurs.

In order to study this asymmetric situation we write a mean-field
equation valid for every value of $h\in[0,1]$. Starting from
Eq.~\ref{eq:midfield}, we just need to specify how the terms
$p^{\pm}_t(h)$ depend on $h$ (see section \ref{subsec:meanfield}),
\begin{align}
  p_t^+(h) &=\frac{M-\left<\Omega_t\right>}{2M} \nonumber\\
  p_t^-(h) &=\frac{h(1-h)}{h^2+(1-h)^2}\frac{\left<\Omega_t\right>}{M} \ .
\end{align}
The resulting mean-field equation can be integrated in the continuum
limit as in the symmetric case $h=1/2$, yielding the dynamics of the
average value of the order parameter. The critical value $\varphi_c$
on the asymmetry $h$ is obtained again by imposing the segregation
regime conditions $\Delta\left<\Omega_t\right>=0$ and
$\left<\Omega_t\right>=M$. Solving for $\varphi$ gives
\begin{equation}
\varphi_c(h)=\frac{h(1-h)}{1-h(1-h)}
\label{eq:critical_h}
\end{equation}
for the critical value. This relation satisfies the red-blue symmetry
$\varphi_c(h)=\varphi_c(1-h)$ with maximum value $\varphi_c(1/2)=1/3$
(as in Eq.~\ref{eq:Omega_SS_phi}) for the symmetric
case. Fig.~\ref{figure:critical_h}b compares the predicted critical
point from Eq.~\ref{eq:critical_h} to simulations of evolved networks
for different values of $h$ Fig.~\ref{figure:critical_h}a.  This
analysis shows that a situation characterized by a minority conviction
favors segregation for lower values of the choice parameter,
indicating that the symmetric situation is the one in which
segregation can be more easily avoided (the situation is analogous to
the miscibility gap for phase segregation in a binary mixture).

The characteristic duration of the transient before a steady state is
reached is also affected by the presence of a minority conviction. The
solution of the mean-field equation gives 
\begin{equation}
  \tau_{\Omega}(h)=\frac{2M\left[ h^2+(1-h)^2 \right]}{1-\varphi} \ ,
\end{equation}
i.e., the characteristic relaxation time will increase for asymmetric
convictions. This time scale is important in cases where the
segregation dynamics competes with the spreading of
consensus~\cite{Holme2006,Durrett2012}.

\subsection{Scale-invariance close to the transition}

\begin{figure}
  \centering \includegraphics[width=0.48\textwidth]{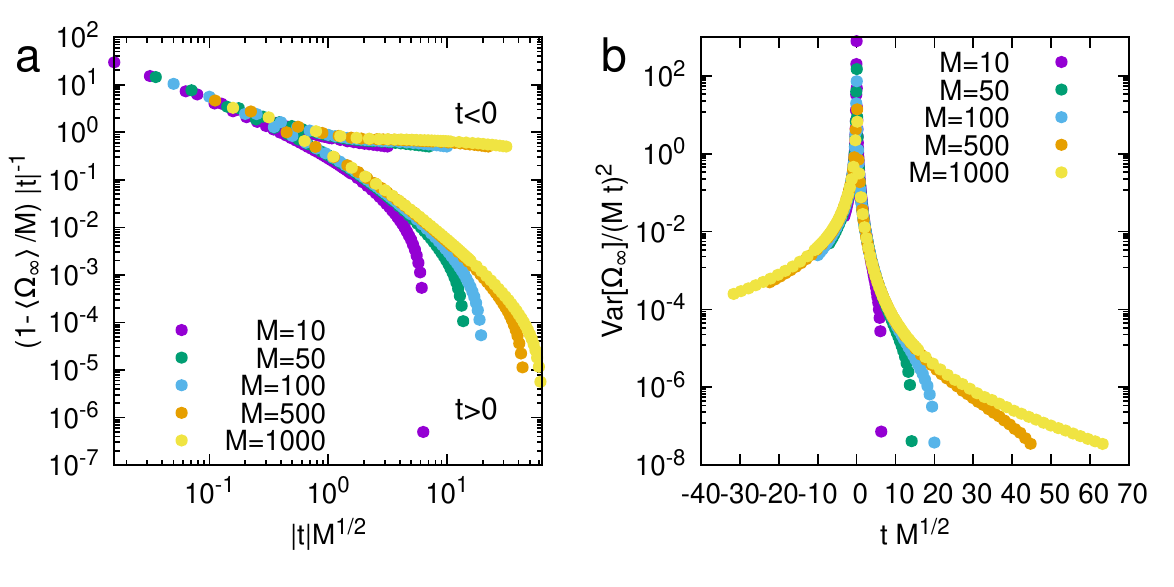}
  \caption{\label{figure:collapse} The fraction of homogeneous edges
    and its variance obey scaling. \textbf{a)} Scaling collapse for
    the fraction of homogenous edges.  \textbf{b)} Scaling collapse
    for the variance. The $x$ and $y$ axes of both plots compare the
    functions predicted by Eqs. \ref{eq:scaling_omega} and
    \ref{eq:scaling_varomega}. The symbols correspond to data points
    from simulations at different network size above and below the
    segragation transition point.}
\end{figure}

The limit of large system size, $M\rightarrow \infty$, is better
analyzed in terms of a finite-size scaling ansatz, typical of critical
phenomena \cite{Fisher1967,hahne2006critical}.  We define the
normalized choice parameter
\begin{equation}
  t = \frac{\varphi - \varphi_c}{\varphi_c} \ . 
\end{equation}
and the intensive order parameter
\begin{equation}
m = \frac{M-\Omega_{\infty}}{M}
\end{equation}
so that
\begin{equation}
\langle m \rangle= 1-\frac{\langle \Omega_{\infty} \rangle}{M}=
\langle \frac{M-\Omega_{\infty}}{M} \rangle
\end{equation}
and we assume that $\langle m \rangle$, which in principle depends on
both $M$ and $t$ separately, is an homogeneous function of $t$ and a
suitable power of $M$, that is
\begin{equation}
\langle m \rangle =
  |t|^{\beta} \tilde{f}_1(M^{y} t) 
  \label{eq:scaling_omega}
\end{equation}
in the large (small) $M$ ($t$) limit with $M^{y} t$ fixed. $y$ and
$\beta$ are exponents that are expected to be independent of the
microscopic details of the dynamical model, characterizing the
transition point, while $f$ is a scaling function, which might depend
on the model specificities. Since we expect that $m$ is non-zero
(zero) for $t<0$ ($t>0$) the scaling function $f$ should behave
asymptotically as
\begin{equation}
\lim_{x\rightarrow +\infty} \tilde{f}_1(x) = 0, \qquad
\lim_{x\rightarrow -\infty}\tilde{f}_1(x) = \textrm{constant} > 0 
\end{equation}
In order to estimate the two scaling exponents $\beta$ and $y$, we plot
$m |t|^{-\beta}$ versus $M^{y} t$ and determine the exponents so that
the best collapse of the different curves is obtained. Indeed one
should obtain a different curve for each value of $M$ as $t$ varies
and this is what we observe for generic pair $\beta$ and $y$.  However
for $\beta = 1$ and $y = 1/2$ the various curves collapse in a range
of $x \equiv M^{y} t$ that increases as $M$ becomes larger and larger
as Fig.\ref{figure:collapse}, panel (a), shows.

The same analysis leads to the following scaling ansatz for the
variance of $m$ (corresponding to $\text{Var}[\Omega_{\infty}]/M^2)$
in terms of the original extensive order parameter):
\begin{equation}
\text{Var}[m] = t^{2} \tilde{f}_2(M^{1/2} t) 
  \label{eq:scaling_varomega}
\end{equation}
and the corresponding collapse is shown in Fig.\ref{figure:collapse},
panel (b). Both scaling Eqs.(\ref{eq:scaling_omega}) and
(\ref{eq:scaling_varomega}) are captured by the more general scaling
ansatz of the distribution function of $m$
\begin{equation}
P(m, t, M)  = |t|^{-1} \tilde{P}(mt^{-1},\ M^{1/2} t) 
  \label{eq:pdf}
\end{equation}

\subsection{A model with pure intra-specific aversion leads to an
  equivalent segregation threshold behavior.}

Motivated by the literature on segregation models based on aversion
between unlike individuals~\cite{Schelling1971,Henry2011}, we asked
whether the same threshold phenomenon observed in our model could be
present in case of conviction moves that were based purely on aversion
bias.

To this end, we defined a model variant where the conviction move
(with probability $\varphi$) chooses randomly one heterogeneous edge,
between two nodes holding different convictions and rewires it to a
random node.  In this variant, the popularity move (with probability
$1-\varphi$ at each step) remains the same. Under this variant,
Eq.~\eqref{eq:midfield} becomes
\begin{equation}
  \Delta\left<\Omega_{t}(\varphi,h)\right> =
  \underbrace{\vartheta\frac{\varphi}{2}}_{\text{conv. move}} +
  \underbrace{(1-\varphi)\left[\vartheta
      p_{t}^{+}(h)-p_{t}^{-}(h)\right]}_{\text{pop. move}}   \ ,
  \label{eq:aversion}
\end{equation}
immediately leading to the expression,
\begin{equation}
\frac{\left< \Omega_{\infty} (\varphi)
  \right>}{M}=\min_{\varphi\in[0;1)} \left\{
  1,\frac{1}{2(1-\varphi)}\right\} \label{eq:Omega_SS_phi2} 
\end{equation}
for the mean fraction of heterogeneous edges. 

\begin{figure}
  \centering \includegraphics[width=0.48\textwidth]{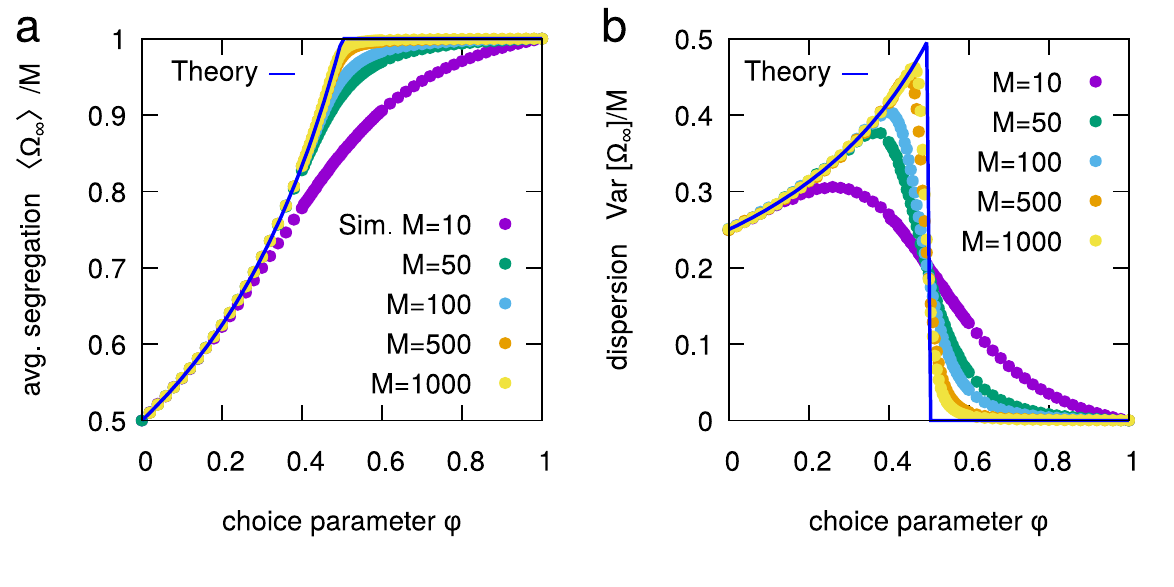}
  \caption{\label{figure:aversion} The sudden transition
      to a segregated state remains in a model with aversion bias
      only. \textbf{a)} Mean order parameter at steady state versus
    the choice parameter $\varphi$ comparing theory (solid line) with
    numerical simulations for different sizes of the network $M$
    (symbols).  This analysis supports a segregation transition for
    $\varphi_c=1/2$ (for $h=1/2$).  \textbf{b)} The dispersion of the
    order parameter (symbols) shows the same behavior as the standard
    model (compare with Fig~\ref{figure:ph_transition}).}
\end{figure}

By setting $\left< \Omega_{\infty} (\varphi) \right>=1$ in
Eq.~\ref{eq:Omega_SS_phi2} and solving for $\varphi$ one finds again
the critical value, which for $h=1/2$ is $\varphi_c=1/2$.
An analogous reasoning
can be followed for solving for the higher moments of the distribution
of $\Omega$.
Fig.~\ref{figure:aversion} shows that direct simulations of the
aversion bias model are fully in line with these theoretical
predictions.  Thus, we conclude that aversion alone is sufficient to
produce a sudden segregation threshold.

\section{Discussion and Conclusions}

Social segregation is ubiquitous in our society, and manifests itself
as fragmentation of social networks at all scales, in countries,
cities, schools, firms, governmental agencies, etc. Its consequences
may lead to a wide range of nefastous phenomena ranging from
inefficient planning to war.
It is driven by diverse and enormously complex sociological, cultural,
environmental and economic dilemmas, which are unlikely to be solved
in the near future.
However, since the pioneering work of
Schelling~\cite{Schelling1969,DallAsta2008,Gauvin2010,Henry2011}
there is increasing agreement that there may be common quantitative
traits in the ``macroscopic'' dynamics of segregation that emerge from
this complexity.
A quantitative understanding of the consequences of such simple
features on the dynamics of a social network may be important to
develop efficient estimators to be used in real-life examples to
detect and prevent segregation phenomena.

The framework developed here shows that complete segregation in a
network setting without any spatial aspects can emerge as a threshold
phenomenon that corresponds to a genuine phase transition. Close to
such transition point, small perturbations of the system can cause
very large rearrangements in the state.
Importantly, we have shown that such transition point is scale
invariant, hence ``universal'' in the statistical physics sense. This
supports the hypothesis that close to this critical point more
detailed descriptions of social interactions are not necessary, since
a wide class of models may behave similarly.

We can also parallel this model with available physical models for the
separation of phases and mixtures.  For example, binary mixtures can
be described in a coarse-grained way as a set of particles of two
kinds filling a cubic lattice, with an energy cost for particles of
one kind sitting next to particles of the other kind. This system
(equivalent to an Ising model) shows a spatial phase separation when
temperature is lowered. Contrary to this case, in our model set on a
network a concept of distance is missing, since all individuals can
potentially interact with any other agent in each move.
However, we can parallel our results to a variant of the above model
where instead of the usual ``local'' fraction of lattice sites
occupied by each kind of particle, we write the free energy in terms
of the parameter used here, i.e., the fraction of homogeneous edges
$e_h = - \Omega/M$.  The energetic term is simply $ -\chi e_h $. In
order to write the entropy, we consider the network as a gas of edges
formed by connecting nodes. We compute the number of ways to assign
$\Omega$ edges out of $M$, considering that each edge is spurious if
two colors of the same kind are selected.  The resulting free energy
is $ \beta F = e_h \log(e_h)+ (1-e_h) \log(1-e_h) - e_h \chi
$. Minimizing this free energy and comparing with the equations
governing our model shows that they are different, and our model
cannot be reconducted to this simple case. The question remains open
on whether there is a simple equilibrium model recapitulating the
phase-separation behavior shown by our segregation model.

Segregation in social networks may be driven by both homophyly (the
choice of social interactions with like individuals) and aversion.
These ingredients are mixed in different proportion in the existing
literature. Our basic model contains both, since in the
conviction-based rewirings interactions between dissimilar partners are
rewired in favor of homogeneous ones.
Schelling's model~\cite{Schelling1971} shows that aversion from
dissimilar network partners alone, coupled with a random selection of
new partners, may be sufficient to induce segregation. Our analysis of
a model variant where the conviction-based rewiring is based on pure
aversion supports this conclusion. Indeed, this variant shows the same
type of threshold phenomenon, in full quantitative agreement with the
main model. The (expected) quantitative change is that in the case of
pure aversion the transition point is shifted to higher values of the
choice parameter $\varphi$, compared to the case where both aversion and
homophyly are in place.

Overall, our analysis supports the conclusion that whether
conviction-based rewiring is based on aversion or homophyly is not a
key ingredient for the existence of a segregation threshold.
Instead, the important feature to determine a threshold phenomenon for
segregation is that the the conviction-based rewiring of the network
(based on aversion or homophyly, or both) occurs on the same time
scale of the popularity-based rewirings (i.e. the establishment of
social interactions that are non-discriminant). In the alternative
scenario in which, e.g., each kind of rewiring occurs proportionally
to the number of extant interactions, segregation occurs smoothly. In
such situation, at all levels of the bias in establishing interactions
(quantified by the choice parameter $\varphi$) the network maintains a
finite fraction of interactions between dissimilar individuals.

\begin{acknowledgments}
  The authors would like to thank Mirta Galesic for useful feedback,
  and Alessandro Civeriati, Andrea Possenti and Sara Cerioli for
  preliminary work on this project.
\end{acknowledgments}

\bibliography{biblio}

\newpage

\appendix

\section{Analytical calculations}
This section presents in further detail the two different methods used to
derive the analytic expressions for the cumulants of the order
parameter (namely equations \ref{eq:Omega_SS_phi} and
\ref{eq:Omega2_SS_phi}).
\subsection{Mean-field approach}
\label{subsec:meanfield}
As previously explained, the mean-field approach consists in
quantifying the average variation of the order parameter at every step
of the dynamics, which resulted in equation \ref{eq:midfield}. The
meaning of the terms of such equation have already been discussed,
here we will present the more general derivation of the contributions
$p_t^\pm(h)$ for every $h\in[0,1]$, which will yield the more general
solution of equation \ref{eq:Omega_time} for different densities of
colored nodes.

The terms $p_t^\pm(h)$ represent the probabilities of, respectively,
increasing and decreasing the order parameter $\Omega$ when a
popularity move is performed:
\begin{align}
p_t^+(h) &=	\mathrm{Prob} \left[ e_t(rb) \to e_t(rr) \right] + 
		  	\mathrm{Prob} \left[ e_t(br) \to e_t(bb) \right] \nonumber\\
p_t^-(h) &= \mathrm{Prob} \left[ e_t(rr) \to e_t(rb) \right] + 
		  	\mathrm{Prob} \left[ e_t(bb) \to e_t(br) \right]
\end{align}
which are found to be 
\begin{align}
p_t^+(h) &=	\frac{M-\langle \Omega_t(\varphi,h) \rangle}{2M} \nonumber\\
p_t^-(h) &= \frac{\langle \Omega_t(\varphi,h) \rangle}{M}\frac{h(1-h)}{h^2+(1-h)^2} \ .
\end{align}
By substituting these coefficients in equation \ref{eq:Omega_time} and
taking the continuous-time limit we obtain the following differential
equation,
\begin{footnotesize}
\begin{equation}
\partial_t \langle \Omega_t(\varphi,h)
\rangle=\vartheta\frac{1+\varphi}{2}-\\(1-\varphi)\frac{2h(1-h)(1-\vartheta)+\vartheta}{2\left(h^{2}+(1-h)^{2}\right)}\frac{\langle
  \Omega_t(\varphi,h) \rangle}{M} \ ,
\end{equation}
\end{footnotesize}
which can be explicitly integrated in time (for $\varphi\neq1$),
yielding
\begin{footnotesize}
\begin{multline}\label{eq:Omega_phi_h_time}
\frac{\langle \Omega_t(\varphi,h) \rangle}{M} =
\left[ 1+\vartheta \left( \frac{\langle \Omega_0(\varphi,h) \rangle}{M}-1 \right) - \vartheta \frac{1+\varphi}{2} \alpha(\varphi,h) \right] \cdot \\
 \cdot e^{-\frac{t}{\alpha(\varphi,h)}} + \vartheta
 \frac{1+\varphi}{2} \alpha(\varphi,h) \ ,
\end{multline}
\end{footnotesize}
where the initial condition is
\begin{equation}
\frac{\langle \Omega_0(\varphi,h) \rangle}{M}= e_0(rr)+e_0(bb)= h^2+(1-h)^2
\end{equation}
and the coefficient $\alpha$ is
\begin{equation}
\alpha(\varphi,h)=\frac{2(h^2+(1-h)^2)}{(1-\varphi)(2h(1-h)(1-\vartheta)+\vartheta)}
\ .
\end{equation}
If we evaluate this coefficient in the unsegregated phase (where
$\vartheta\equiv1$), we obtain the characteristic time of the
transient phase, which is
\begin{equation}
\tau(\varphi,h)=\frac{2(h^2+(1-h)^2)}{1-\varphi}
\end{equation}
Taking the limit $t\to\infty$ of equation \ref{eq:Omega_phi_h_time}
yields the steady-state solution of the order parameter, which for
every $\varphi\in[0,1)$ and $h\in[0,1]$ is,
\begin{equation}
\frac{\langle \Omega_t(\varphi,h) \rangle}{M}=\min
\left\{1,\frac{1+\varphi}{1-\varphi}\left(h^2+(1-h)^2\right)\right\} \ .
\end{equation}
%
%
%
%
%
Fig.~\ref{figure:critical_h} shows the phase diagram for
$\langle \Omega_t \rangle$, which is in agreement with the fact that
the critical value of the choice parameter $\varphi_c$ becomes lower
as we move away from the symmetric nodes density given by $h=1/2$
(discussed in section \ref{sec:min_conv}).

\subsection{Master equation and moment-generating function approach}
\label{subsec:ME_FMGF}

This section treats in further detail the derivation of a generic
factorial moment of the order parameter $\Omega$. Substituting the
rates \ref{eq:ME_rates} in the master equation \ref{eq:ME} one gets,
\begin{align}\label{eq:ME_full}
\partial_t P_t(\Omega) &=P_t(\Omega-1)\left[
                         \varphi+(1-\varphi)\frac{M-\Omega+1}{2M}
                         \right] + \nonumber \\ 
&+ P_t(\Omega+1)(1-\varphi)\frac{\Omega+1}{2M}-P_t(\Omega)\frac{1+\varphi}{2}
\end{align}
In order to find a differential equation for the FMGF \ref{eq:FGMF} we
first multiply by $s^{\Omega}$ both sides of equation
\ref{eq:ME_full}, and then we sum over the order parameter $\Omega$
itself.  The probabilities $P_t(\Omega)$ are obviously defined only
for $\Omega\in[0,M]$, so we need to explicitly set
$P_t(\Omega)\equiv0$ when $\Omega$ is outside that range.  This
notation has a practical advantage that allows us to extend the
summation over $\Omega$ from the range $[0,M]$ to the range
$[-1,M+1]$. This frees from border-term issues when re-indexing the
summation for the terms on the right side.  To evaluate the
contribution with the $P_t(\Omega-1)$ coefficient, we set
$\Omega'=\Omega-1$ and obtain
\begin{multline}
\sum_{\Omega'=-2}^{M}s^{\Omega'+1} P_t(\Omega') \left[ \frac{1+\varphi}{2} -\frac{\Omega'}{2M}\right] = \\
=\left[ s\frac{1+\varphi}{2}-\frac{1-\varphi}{2M}s^2\partial_s \right]
G(s,t) \ , 
\end{multline}
where we introduced a derivative in $s$ in order to eliminate the
multiplicative $\Omega'$ in the summation. The same trick can be used
for the $P_t(\Omega+1)$ term (this time we set $\Omega'=\Omega+1$):
\begin{equation}
\sum_{\Omega'=0}^{M+2}s^{\Omega'-1}(1-\varphi)P(\Omega')\frac{\Omega'}{M} = 
\frac{1-\varphi}{2M}\partial_s G(s,t)
\end{equation}
Finally, the $P_t(\Omega)$ term does not require any re-indexing and
immediately yields $G(s,t)(1+\varphi)/2$. Putting all the pieces
together we finally find the desired equation \ref{eq:FGMF_dyn} for
the dynamics of the FMGF.

Equation \ref{eq:FGMF_dyn} is a partial differential equation that
contains derivatives both in $s$ and $t$. Since we are only interested
in finding the moments of the equation, we can avoid solving it
explicitly: if we evaluate $\partial_s^n[\partial_t G(s,t)|_{s=1}]$
for every $n\in\mathbb{N}$ we obtain a closed system of time-only
differential equations for the dynamics of the moments. In fact we can
easily see that
\begin{equation}
\partial_s^n G(s,t)|_{s=1} = \langle \frac{\Omega!}{(\Omega-n)!} \rangle
\end{equation}
For $n=1$, we are evaluating the first factorial moment, which
coincides with the average. A straightforward calculation shows that
we obtain precisely equation \ref{eq:dt_Omega} (in the unsegregated
phase with $\vartheta\equiv1$).  For $n=2$, we find the equation of
the second factorial moment
$\langle \Omega(\Omega-1) \rangle = \langle \Omega^2 \rangle -\langle
\Omega \rangle$, which reads
\begin{equation}
\partial_t\langle \Omega^2 \rangle -\partial_t \langle \Omega \rangle = 
-2\frac{1-\varphi}{M}\langle \Omega^2 \rangle + 
\left( 1+\varphi +\frac{1-\varphi}{M} \right) \langle \Omega \rangle
\end{equation}
By evaluating the steady-state solution
($\partial_t\langle \Omega^2 \rangle = \partial_t \langle \Omega
\rangle=0$) of this equation and substituting the steady-state form of
$\langle \Omega \rangle$, we find the steady-state equation of
$\langle \Omega^2 \rangle$, which in turn gives us the variance
\begin{equation}
\mathrm{Var}\left[ \Omega \right] = \langle \Omega^2 \rangle - \langle
\Omega \rangle^2 = \frac{1+\varphi}{4(1-\varphi)} \ .
\end{equation}
This equation coincides with the one presented in equation
\ref{eq:Omega2_SS_phi} (in the unsegregated phase).

\subsection{Full Stationary Solution}
\label{subsec:ME_STATSOL}

Starting from the Master Equation (\ref{eq:ME_full}) we can write the
full equation for the Generating Function $G(s,t)$ 
\begin{equation}\label{fullEqG}
\partial _tG(s,t)=a G(s,t) (s-1)+b \partial _sG(s,t)\left(1-s^2\right)
\ ,
\end{equation}
where $a=\frac{1+\varphi }{2}$ and $b=\frac{1-\varphi }{2 M}$; $M$ is
the total number of links. We assume the initial condition
($P(\Omega,t=0)=\delta _{\Omega -M/2}$ and thus we have
$G(s,0)=s^{M/2}$. Additionally, the normalisation condition fixes
$G(1)=1$.

The stationary solution for Eq. (\ref{fullEqG}) is simple to find by
solving directly the PDE, and leads to
\begin{equation}\label{StatG}
G(s)= \left(\frac{1+ s}{2}\right)^{a/b} \ .
\end{equation}

In order to solve the full transient of the PDE (\ref{fullEqG}) we use
the so-called method of characteristics. Setting
$f(s)=-b\left(1-s^2\right)$, then Eq. (\ref{fullEqG}) corresponds to
the following system of differential equations:
\begin{eqnarray}
  \label{CMode1}\dot{s}(t)&=&f(s)\\
  \label{CMode2}\frac{d}{dt}G(s(t),t)&=&a(s(t)-1)G(s(t),t) \ .
\end{eqnarray}
Eq. (\ref{CMode1}) leads to the integral equation
$-\int_{s(0)}^s \frac{dz}{1-z^2} \, dz=\int _0^tb dt$ where $s$ it
evaluated at a final time $t$, i,e, $s(t)=s$. Solving this equation
leads to
\begin{equation}
s=\frac{\text{Cosh}(b t) s(0)-\text{Sinh}(b t)}{\text{Cosh}(b t)-s(0) \text{Sinh}(b t)}
\end{equation}
and
\begin{equation}
s(0)=\frac{s \text{Cosh}(b t)+\text{Sinh}( b t)}{\text{Cosh}( b t)+s
  \text{Sinh}( b t)} \ .
\end{equation}
Finally, performing the integral $\int _{s(0)}^s a(s(\tau)-1)d\tau$ we
find
\begin{multline}
G(s,t)=e^{-a t} (\text{Cosh}( b t)+s \text{Sinh}( b t))^{a/b} \cdot \\
\cdot \left(\frac{s \text{Cosh}( b t)+\text{Sinh}(b t)}{\text{Cosh}( b
    t)+s \text{Sinh}( b t)}\right)^{M/2} \ . 
\end{multline}

In the limit t$\to \infty $, this expression gives the stationary
solution Eq.~(\ref{StatG}). Expanding this in series around $s=0$, and
matching term by term, one can find the transient solution
$P(\Omega,t)$. In fact, we have that
$G(s,t)=P(0,t)+sP(1,t)+...+s^M P(M,t)$ and $G(0,t)=P(0,t)$.  Expanding
the steady state solution of $G(s)$ in series around $s=0$, we obtain
$G(s)=\sum_{\Omega =0}^M \frac{\partial_s^{\Omega}G(s)_{| s=0}}{\Omega
  !}s^{\Omega}$ leading to Eq. (\ref{Pstaz}) in the main text. We
highlight that Eq. (\ref{Pstaz}) only holds for
$\varphi \in [0, \varphi_c)$.


\end{document}